\documentclass[11pt]{raa}

\usepackage{graphicx,times}
\usepackage{natbib}
\usepackage{amssymb,amsmath}
\usepackage[pagebackref=true]{hyperref}
\bibpunct{(}{)}{;}{a}{}{,}

\begin{document}

\title{Cosmological constraints on neutrino masses in light of JWST red and massive candidate galaxies}

\volnopage{ {\bf 20XX} Vol.\ {\bf X} No. {\bf XX}, 000--000}
   
\setcounter{page}{1}

\author{Jian-Qi Liu\inst{1}, Zhi-Qi Huang\inst{*1,2}, Yan Su\inst{1}}

\institute{
    School of Physics and Astronomy, Sun Yat-sen University, 2 Daxue Road, Tangjia, Zhuhai, 519082, China\\
    \and
    CSST Science Center for the Guangdong-Hong Kong-Macau Greater Bay Area, Zhuhai, 519082, China; {\it huangzhq25@mail.sysu.edu.cn}\\
    \vs \no
    {\small Received 20XX Month Day; accepted 20XX Month Day}
}

\abstract{The overabundance of the red and massive candidate galaxies observed by the James Webb Space Telescope (JWST) implies efficient structure formation or large star formation efficiency at high redshift $z\sim 10$. In the scenario of a low or moderate star formation efficiency, because massive neutrinos tend to suppress the growth of structure of the universe, the JWST observation tightens the upper bound of the neutrino masses. Assuming $\Lambda$ cold dark matter cosmology and a star formation efficiency $ \in [0.05, 0.3]$ (flat prior), we perform joint analyses of Planck+JWST and Planck+BAO+JWST, and obtain improved constraints $\sum m_\nu < 0.196\,\mathrm{eV}$ and $\sum m_\nu < 0.111\,\mathrm{eV}$ at 95\% confidence level, respectively. Based on the above assumptions, the inverted mass ordering, which implies $\sum m_\nu\geq 0.1\mathrm{eV}$, is excluded by Planck+BAO+JWST at 92.7\% confidence level. 
}

\authorrunning{J.-Q. Liu et al. }            
\titlerunning{Cosmological constraints on neutrino masses in light of JWST}  

\maketitle

\section{Introduction}
\label{sect:intro} 

The standard hot big bang cosmology predicts a cosmic neutrino background (C$\nu$B), which decoupled from the thermal bath in the early universe at a temperature $\sim \mathrm{MeV}$. The subsequently redshifted momenta of C$\nu$B follow an ultrarelativistic Fermi-Dirac distribution with negligible chemical potential and an effective temperature $1.95(1+z)\,\mathrm{K}$, where $z$ denotes the cosmological redshift. Although direct detection of C$\nu$B is yet unrealistic, the existence of C$\nu$B has been indirectly confirmed by the observations of primordial abundances of light elements and the cosmic microwave background (CMB). 

The difference between the squares of the three mass eigenstates has been measured in neutrino flavor oscillation experiments~\citep{fukuda1998evidence, abe2014observation}. These experiments constrain the sum of masses of three neutrino species, $\sum m_\nu$, to be $\geq 0.06\,\mathrm{eV}$ for the normal mass ordering $m_1\sim m_2 \ll m_3$, and $\geq 0.10\,\mathrm{eV}$ for the inverted mass ordering $m_3\ll m_1 \approx m_2$. Distinguishing between the two mass-ordering scenarios, and hence accurate measurements of the sum of neutrino masses, are important for understanding the origin of neutrino masses.

At $z\approx 1100$ where primary CMB anisotropies are generated, the background temperature is $T\approx 3000\,\mathrm{K}\approx 0.3\,\mathrm{eV}$. If $\sum m_\nu \gtrsim 0.3\,\mathrm{eV}$, neutrino masses will have an impact on the primary CMB via the early integrated Sachs Wolfe effect. In the late universe, massive neutrinos act as a hot dark matter component, which tends to suppress the growth of large scale structure at small scales ($\lesssim 100\,\mathrm{Mpc}$) and alter the CMB lensing effect. The third generation CMB experiment Planck mission constrained the total neutrino masses to be $\sum m_\nu < 0.241\,\mathrm{eV}$ (TTTEEE + lowE + lensing) at 95\% confidence level (CL) in the standard $\Lambda$ cold dark matter ($\Lambda$CDM) model~\citep{aghanim2020planck}. Further including baryon acoustic oscillation (BAO) data , which breaks the degeneracy between neutrino masses and the background expansion history of the universe, pushes the upper limit of neutrino masses to $\sum m_\nu < 0.121\,\mathrm{eV}$ (95\% CL). This upper bound already puts inverted mass ordering under pressure (excluded at 90.2\% CL).

The standard cosmological structure and galaxy formation is recently challenged by red and massive candidate galaxies from James Webb Space Telescope (JWST). \citealt{labbe2023population} found six candidate massive galaxies (stellar mass $10^{10}$ solar masses) at $7.4\le z \le 9.1$. This finding suggests that either the star formation efficiency (SFE) significantly exceeds its typical values in low-redshift galaxies, or the halo mass function is about $2\sigma$ higher than the prediction of the standard $\Lambda$CDM cosmology~\citep{boylan2023stress, lovell2023extreme, haslbauer2022has, qin2023implications, wang2023quantifying}. Although at current stage a large SFE cannot be theoretically excluded~\citep{zhang2022massive, qin2023implications}, a large SFE at $z\gtrsim 7$ tends to enhance the UV radiation from galaxies, causing a faster reionization process that is in tension with observation~\citep{goto2021silverrush, wold2022large, lin2023implications}. For this reason, we shall take the more widely accepted conservative assumption of $\mathrm{SFE}\lesssim 0.1$, and discuss the cosmological implications. Because massive neutrinos tend to suppress the number of massive halos in which massive galaxies are born,  a large $\sum m_\nu$ increases the tension between $\Lambda$CDM model and JWST data. Thus, adding JWST data to the cosmological analysis, as the present work aims to do, helps to tighten the upper bound of neutrino masses.

This paper is organized as follows. Section~\ref{sect:method} constructs the likelihood of massive galaxy counting and describes the numerical tool for the likelihood evaluation. In section~\ref{sect:results} we apply the likelihood to the \citealt{labbe2023population} data and obtain a Planck+JWST and Planck+BAO+JWST joint constraints on $\sum m_\nu$. Section~\ref{sect:conc} discusses and concludes.

\section{Likelihood of massive galaxy counting} \label{sect:method}

The aim of this section is to extend the likelihood of massive galaxy counting in \citealt{wang2023quantifying}, which we sketch below, to massive neutrino models. 

To assess the impact of massive neutrinos on galaxy formation, we adopt the extended Press-Schechter ellipsoidal collapse model(\citealt{press1974formation,sheth2001ellipsoidal,sheth2002excursion}) to compute the expected abundance of dark matter halos. It predicts the halo mass function, which is the comoving halo number density per mass interval, to take the form
\begin{equation}
\frac{\mathrm{d}N}{\mathrm{d}M} = \frac{\Bar{\rho}}{M^2}\frac{\mathrm{d}\ln \nu}{\mathrm{d}\ln M}f(\nu), \label{eq:haloMF}
\end{equation}
Here, $\Bar{\rho}$ is the average density of the background and $\nu=\delta_c/\sigma(M,z)$, where $\delta_c=1.686$ corresponds to the critical linear overdensity and $\sigma(M,z)$ is the mass fluctuation at scale $R=(3M/4\pi\Bar{\rho})^\frac{1}{3}$. The simulation-calibrated  $f(\nu)$ factor is given by
\begin{equation}
   f(\nu) = 2A (\frac{1}{\nu'^{2q}}+1) \frac{\nu'^{2}}{2\pi} e^{\frac{-\nu'^{2}}{2}}, 
\end{equation}
where $\nu'=\sqrt{a}\nu$, $a=0.707$, $A=0.322$ and $q=0.3$~\citep{sheth1999large}. For massive halos we assume that the fraction of baryonic mass in a massive dark matter halo is $f_b \equiv \Omega_b/\Omega_m$, where $\Omega_b$ and $\Omega_m$ are the parameters of baryon and matter abundance. The stellar mass $M_*$ is then connected to the halo mass $M_{\rm halo}$ via $M_* = \epsilon f_b M_{\rm halo}$, where $\epsilon$ is the SFE. This relation allows to convert a stellar-mass threshold $M_{\rm *,cut}$ to a halo-mass threshold $M_{\rm halo, cut} = \frac{M_{\rm *,cut}}{\epsilon f_b}$. Assuming that each massive halo has a massive central galaxy, we can write the expected number of massive galaxies above the stellar-mass threshold and in a selected comoving volume as
 \begin{equation}
    \langle N_{\rm th} \rangle=4\pi f_{\rm sky}\int_{M_{\rm halo, cut}}^{\infty}\mathrm{d}M \int_{z_{\min}}^{z_{\max}} \frac{\mathrm{d}n}{\mathrm{d}M} \frac{\mathrm{d}V}{\mathrm{d}z\mathrm{d}\Omega} \,\mathrm{d}z, \label{eq:Nth}
\end{equation}
where the selected comoving volume is defined by the redshift interval $[z_{\min}, z_{\max}]$ and the sky fraction $f_{\rm sky}$. Following \citealt{wang2023quantifying} we take $z_{\min}=7$ and $z_{\max} = 10$. The survey area is $38\,\mathrm{arcmin}^2$, which gives $f_{\rm sky}=2.56\times 10^{-7}$. The comoving volume per redshift interval per solid angle, $\frac{\mathrm{d}V}{\mathrm{d}z\mathrm{d}\Omega}$, is specified by the cosmology.  For spatially flat $\Lambda$CDM model, to a very good approximation, we have 
\begin{equation}
 \frac{\mathrm{d}V}{\mathrm{d}z\mathrm{d}\Omega} = \frac{c^3}{H_0^3\sqrt{\Omega_m(1+z)^3 + 1-\Omega_m}} \left[\int_0^z \frac{dz'}{\sqrt{\Omega_m(1+z')^3 + 1-\Omega_m}}\right]^2, \label{eq:dV}
\end{equation}
where $c$, $\Omega_m$ and $H_0$ are the speed of light,  the total matter (CDM + baryon + neutrinos) abundance parameter and the Hubble constant, respectively. 

\begin{figure}
    \centering
    \includegraphics[width=0.75\textwidth]{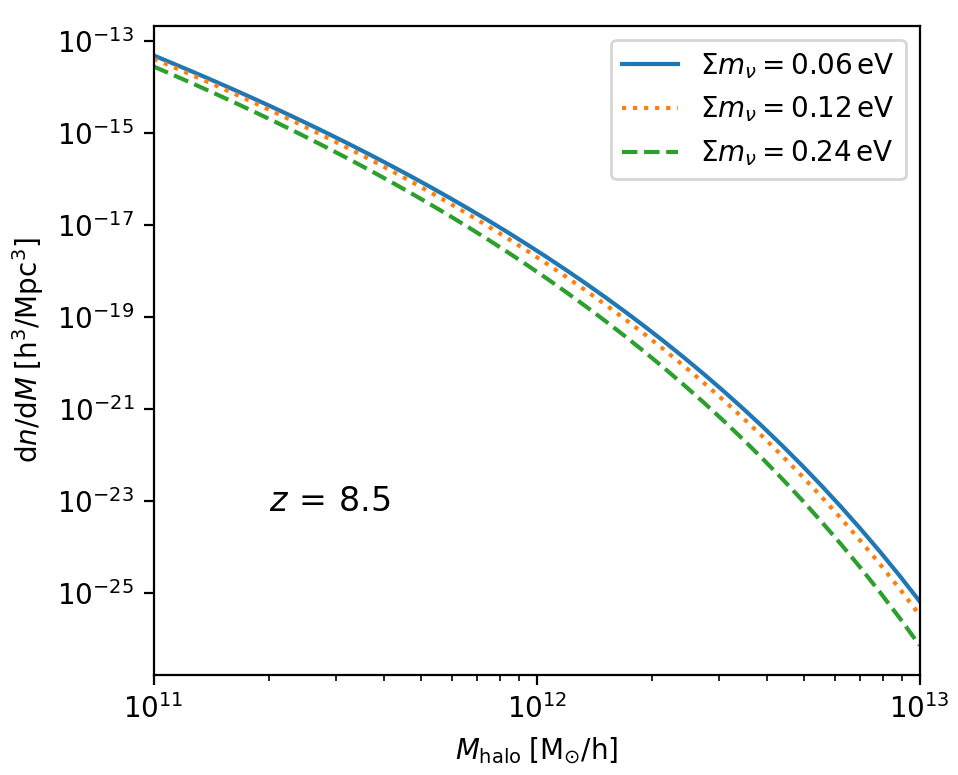}
    \caption{From top to bottom are the halo mass functions at $z=8.5$ for $\sum m_\nu = 0.06\,\mathrm{eV}$, $0.12\,\mathrm{eV}$, and $0.24\,\mathrm{eV}$, respectively. The other cosmological parameters $(\Omega_m, \Omega_b, H_0, A_s, n_s)$ are fixed at Planck best-fit values. }
    \label{fig:haloMF}
\end{figure}

At $z\lesssim 10$, massive neutrinos become non-relativistic and contribute a small ($\sim 10^{-3}$) fraction to the total energy budget of the universe. However, the correction to equation~\eqref{eq:dV} due to neutrino masses, namely the volume effect, is still negligible for our analysis. The major impact of neutrino masses is on the matter power spectrum, integration of which gives $\sigma(M, z)$ in equation~\eqref{eq:haloMF}. Figure~\ref{fig:haloMF} shows the dependence of halo mass function, evaluated at $z=8.5$ (the center of the redshift bin used in the likelihood), on the sum of neutrino masses.
       
The stellar mass and the redshift uncertainties propagate to the uncertainty in $N_{\rm obs}$, the number of  observed galaxies above the stellar mass threshold and within the selected volume. Meanwhile, due to cosmic variance and Poisson shot noise, the theoretical prediction of $N_{\rm obs}$ (denoted as $N_{\rm th}$) is also probabilistic. The likelihood of the theory is then characterized by the probability of finding $N_{\rm obs} \le N_{\rm th}$, which is given by
\begin{equation}
\mathrm{P}\left( N_{\rm obs} \leq N_{\rm th}\right) = \sum_{N_{\rm obs}=0}^\infty \mathrm{P}\left(N_{\rm obs}\right)\sum_{N_{\rm th}=N_{\rm obs}}^\infty \mathrm{P}\left(N_{\rm th}\right). \label{eq:like}
\end{equation}
The distribution function $\mathrm{P}\left(N_{\rm th}\right)$ is based on the cosmological model and the star formation efficiency. For the purpose of using rare-object statistics, we are only interested in the $\langle N_{\rm th} \rangle \lesssim O(1)$ case, where $N_{\rm th}$ approximately follows a Poisson distribution. The expectation-value parameter of the Poisson distribution, denoted as $\lambda$, has an uncertainty due to cosmic variance~\citep{trenti2008cosmic}. Marginalizing over the cosmic variance of $\lambda$ that is denoted as $\sigma_\lambda$, we obtain the distribution function of $N_{\rm th}$, 
\begin{equation}
    \mathrm{P}\left(N_{\rm th}\right)=\int_{-\infty}^{\infty} \frac{1}{\sqrt{2\pi}\sigma_{\lambda}} e^{-\frac{(\lambda-\left \langle N_{\rm th} \right \rangle)^2}{2\sigma^2_\lambda}} e^{-\lambda} \frac{\lambda^{N_{\rm th}}}{N_{\rm th}!} \mathrm{d}\lambda.
\end{equation}
The distribution function $\mathrm{P}\left(N_{\rm obs}\right)$ is obtained by randomly sampling stellar masses and redshifts of the candidate galaxies with the summary statistics given in \citealt{labbe2023population}, and counting the number of galaxies in the selected volume. The precise values of $\mathrm{P}\left(N_{\rm obs}\right)$ rely on the functional form of the systematic errors of the stellar masses and redshifts of the galaxy candidates. We adopt the triangular distribution function, which is shown to be a more conservative estimation than smooth distributions, e.g., the skew-normal distribution. See \citealt{wang2023quantifying} for more detailed description of the algorithm and its theoretical explanation.

\begin{figure}
\centering
\includegraphics[width=0.46\textwidth]{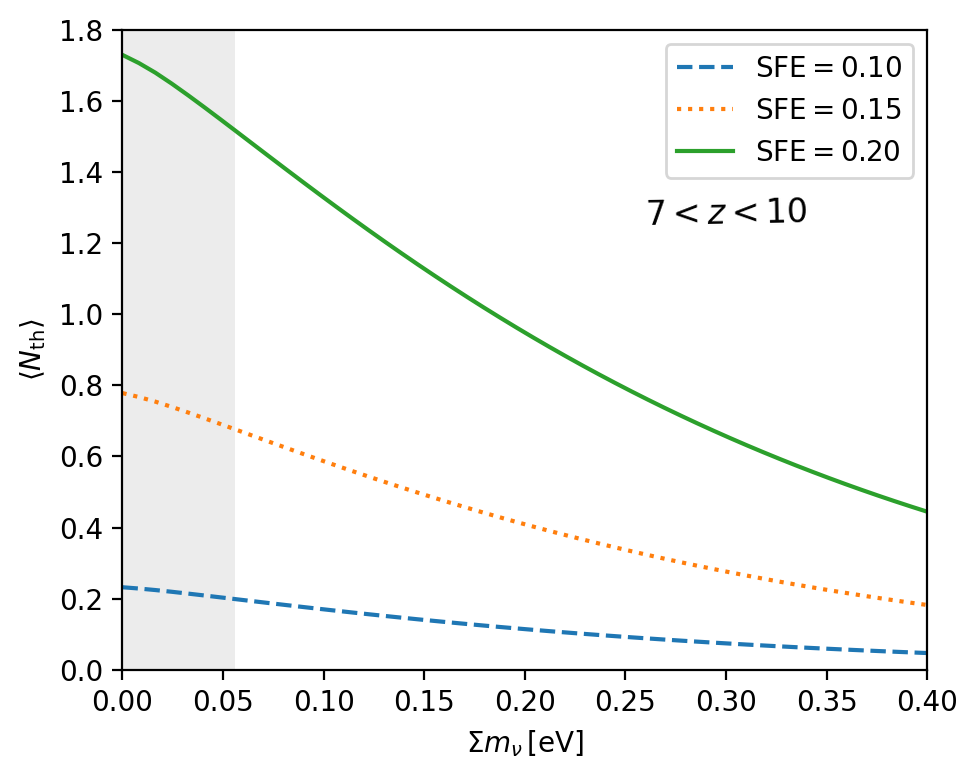}%
\includegraphics[width=0.46\textwidth]{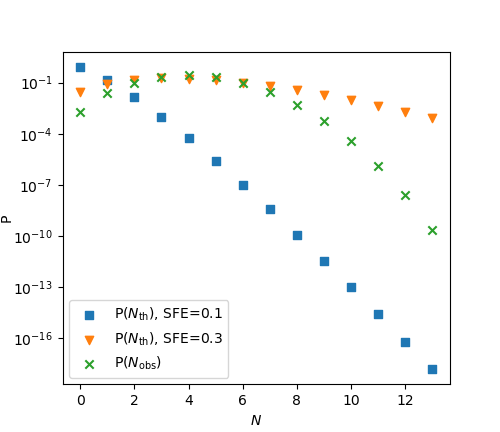}
 \caption{Left panel: dependence of $\langle N_{\rm th}\rangle$ on $\sum m_\nu$ and SFE; other cosmological parameters $(\Omega_m, \Omega_b, h, A_s, n_s)$ are fixed at the Planck best-fit values. Right panel:  $\mathrm{P}\left( N_{\rm obs}\right)$ from the 13 red and massive candidate galaxies in \citealt{labbe2023population} and $\mathrm{P}\left(N_{\rm th}\right)$ from the Planck best-fit $\Lambda$CDM cosmology.} \label{fig:nbar}
\end{figure}

We integrate the cosmological linear perturbation Boltzmann solver CAMB~\citep{howlett2012cmb,lewis2000efficient} into the $\mathrm{P}\left( N_{\rm obs} \leq N_{\rm th}\right)$ calculator in \citealt{wang2023quantifying}.  As shown in the left panel of Figure~\ref{fig:nbar}, the expected number of massive galaxies $\langle N_{\rm th}\rangle$ decreases with the sum of neutrino masses, while the other cosmological parameters $(\Omega_m=0.3158, \Omega_b=0.04939, h=0.6732, A_s=2.101\times 10^{-9}, n_s=0.9661)$ are fixed at the Planck best-fit values~\citep{aghanim2020planck}. Here the reduced Hubble constant $h$ is defined by $H_0=100h\,\mathrm{km/s/Mpc}$; $\Omega_m$ and $\Omega_b$ are the total matter (CDM + baryon + neutrinos) and baryon abundances; $A_s$ and $n_s$ are the amplitude and spectral index of the power spectrum of the primordial comoving curvature perturbations. In the right panel of Figure~\ref{fig:nbar} we show the derived data, namely $\mathrm{P}\left(N_{\rm obs}\right)$ for the 13 candidate galaxies from \citealt{labbe2023population}, and the theoretical $\mathrm{P}\left(N_{\rm th}\right)$ for two representative values $\mathrm{SFE}=0.1$ and $\mathrm{SFE}=0.3$, respectively. For the case of $\mathrm{SFE}=0.1$, there is a tension betwen the theory and the data, as the theory prefers a smaller number of massive galaxies ($N < 2$) than what the data indicates ($N\gtrsim 2$). For $\mathrm{SFE}=0.3$, the theory is well consistent with the data as $\mathrm{P}\left(N_{\rm th}\right)$ is comparable or greater than $\mathrm{P}\left(N_{\rm obs}\right)$ for all $N$'s. 

\section{Results} \label{sect:results}

For the given set of observed galaxy candidates and selected volume ($7\le z \le 10$ and survey area $38\,\mathrm{arcmin}^2$), the cosmology-dependent $\mathrm{P}\left( N_{\rm obs} \leq N_{\rm th}\right)$ in equation~\eqref{eq:like} gives a likelihood of cosmological parameters and SFE. For more detailed information of the data set, such as the redshifts and stellar masses of the candidate galaxies, we refer the readers to \citealt{wang2023quantifying}. 

Due to the degeneracy between cosmological parameters, JWST data alone cannot give a meaningful constraint on $\sum m_\nu$. We therefore analyze Planck+JWST and Planck+BAO+JWST jointly by performing importance sampling with the Planck and Planck+BAO Monte Carlo Markov chains from the Planck Legacy Archive (\url{https://pla.esac.esa.int/pla/}). The base parameters, on which flat priors are assumed, are the baryon density $\Omega_bh^2$, the CDM density $\Omega_ch^2$, the reionization optical depth $\tau$, the angular extension of sound horizon on the last scattering surface $\theta$, the sum of neutrino masses $\sum m_\nu$, the primordial power spectrum parameters $\ln A_s$ and $n_s$, and the SFE. Before doing cosmological analysis, we have to assume a prior on SFE. Unless otherwise stated, we focus our study on a low or moderate SFE, namely a flat prior $\mathrm{SFE}\in [0.05, 0.3]$, though other options will also be briefly discussed.

The constraints on some key parameters (base or derived), before and after incorporating JWST likelihood, are shown in Table~\ref{tab:results} for a comparison. The 95\% CL upper bound of $\sum m_\nu$ is improved by 19\% (without BAO) and 8\% (with BAO), respectively. Although the improvement is insignificant, the inclusion of JWST data pushes the upper bound to $\sum m_\nu < 0.111\,\mathrm{eV}$, very close to the $\sum m_\nu > 0.1\,\mathrm{eV}$ limit implied by the inverted mass ordering. Integrating over the posterior distribution of $\sum m_\nu$, we find that the inverted mass ordering is excluded by Planck+BAO+JWST at 92.7\% CL.

\begin{table}
\centering
\caption{Marginalized $1\sigma$ constraints on cosmological parameters; for $\sum m_\nu$ the 95\% CL upper bounds are shown.\label{tab:results}}
    \begin{tabular}{lllll}
    \hline
    \hline
& Planck & Planck+JWST & Planck+BAO & Planck+BAO+JWST \\
\hline
$\sum m_\nu$ [eV] & $<0.241$ & $<0.196$ & $<0.121$ & $<0.111$\\
$\sigma_8$ & $0.807\pm 0.016$ & $0.811\pm 0.014$ & $0.813\pm 0.010$ & $0.815\pm 0.009$ \\
$H_0$ [km/s/Mpc] & $67.1\pm 1.05$ & $67.3\pm 0.92$ & $67.8\pm 0.51$ & $67.9\pm 0.50$ \\
$100\Omega_bh^2$ & $2.237\pm 0.015$ & $2.238\pm 0.015$ & $2.242\pm 0.014$ &  $2.242\pm 0.014$ \\
$\Omega_ch^2$ & $0.1201\pm 0.0013$ & $0.1200\pm 0.0012$ & $0.1193\pm 0.0009$ & $0.1194\pm 0.0009$\\
$100\theta$ & $1.04082\pm 0.00032$ & $1.04090\pm 0.00031$ & $1.04100\pm 0.00031$ & $1.04094\pm 0.00031$\\ 
$\tau$ & $0.0547\pm 0.0076$ & $0.0553\pm 0.0076$  & $0.0553\pm 0.0073$ & $0.0559\pm 0.0073$\\
$\ln\left(10^{10}A_s\right)$ & $3.046\pm 0.015$ & $3.048\pm 0.015$ & $3.045\pm 0.015$ & $3.046\pm 0.015$  \\
$n_s$ & $ 0.9648\pm 0.0042$ & $0.9652\pm 0.0042$ & $0.9666\pm 0.0037$ & $0.9667\pm 0.0037$ \\
    \hline
    \end{tabular}
\end{table}

\begin{figure}
  \centering
  \includegraphics[width=0.46\textwidth]{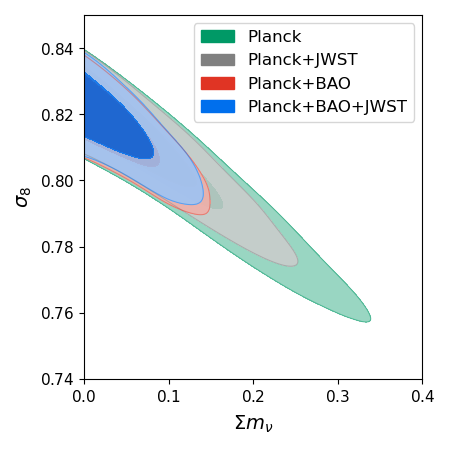}%
  \includegraphics[width=0.46\textwidth]{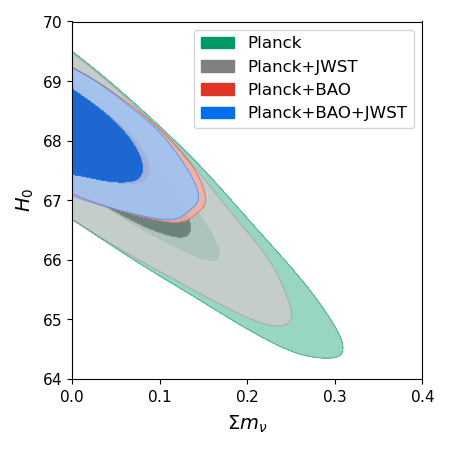}        
  \caption{Marginalized $68\%$ CL and $95\%$ CL constraints on $(\sum m_\nu, \sigma_8)$ (left panel) and $(\sum m_\nu, H_0)$ (right panel). The units are $\mathrm{eV}$ for $\sum m_\nu$ and $\mathrm{km/s/Mpc}$ for $H_0$. The prior on star formation efficiency is $\mathrm{SFE} \in [0.05, 0.3]$.}
  \label{fig:contours}
\end{figure}

The addition of JWST data also has an impact on $\sigma_8$, the root mean square of density fluctuation in a top-hat spherical window with radius $8h^{-1}\mathrm{Mpc}$. This is because the suppression effect due to neutrino masses is degenerate with the amplitude of matter density fluctuations. This degeneracy can be clearly seen in the left panel of Figure~\ref{fig:contours}, which shows the marginalized $1\sigma$ and $2\sigma$ contours of  $\sum m_\nu$ and $\sigma_8$. On the right panel of Figure~\ref{fig:contours} we show the well known geometric degeneracy between $\sum m_\nu$ and $H_0$, which allows the inclusion of BAO (or any other geometric probes at low redshift) significantly improve the CMB-alone constraint on neutrino masses. The figures are made with the publicly available GetDist package~\citep{lewis2019getdist}.

\begin{figure}
  \centering
  \includegraphics[width=0.75\textwidth]{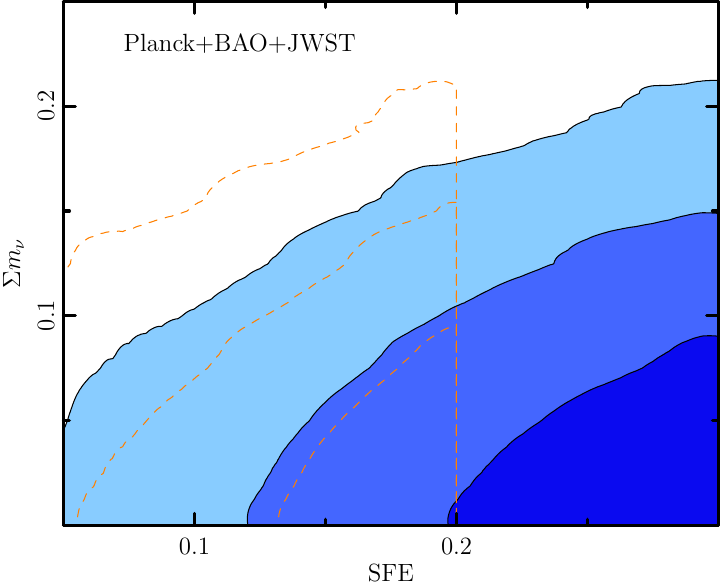}
  \caption{Marginalized $1\sigma$, $2\sigma$ and $3\sigma$ constraints on $\mathrm{SFE}$ and $\sum m_\nu$ with Planck+BAO+JWST. The prior on star formation efficiency is $\mathrm{SFE} \in [0.05, 0.3]$. Orange dashed lines shows how the contours vary when the upper bound of SFE is changed to $0.2$.}
  \label{fig:SFE_mnu}
\end{figure}

The solid and filled contours in Figure~\ref{fig:SFE_mnu} show the marginalized $1\sigma$, $2\sigma$ and $3\sigma$ constraints on SFE and $\sum m_\nu$. Although seemingly the upper bound of $\sum m_\nu$ tightens when SFE is lower, it is {\it not} the case. This is because the likelihood is very sensitive to SFE, and therefore the contours vary significantly when we change the prior on SFE. For example, when the upper bound of SFE is changed to $0.2$, the contours become the orange dashed lines in the figure. When we marginalize over SFE, the upper bound of $\sum m_\nu$ does not change much. See also Table~\ref{tab:SFEprior} for concrete numbers.

\begin{table}
  \centering
  \caption{Dependence of the $95\%$ CL upper bound of $\sum m_\nu$ on the prior on SFE; data sets: planck+BAO+JWST\label{tab:SFEprior}; the last column shows the relative improvement on the $\sum m_\nu$ bound against the planck+BAO (without JWST) result, namely $\sum m_\nu < 0.121\,\mathrm{eV}$ (95\% CL).}
  \begin{tabular}{lll}
    \hline
    \hline
    prior on SFE &  95\% CL upper bound of $\sum m_\nu$  & improvement against planck+BAO\\
    \hline
    SFE$\in [0.05, 0.1]$ &  $\sum m_\nu < 0.118\,\mathrm{eV}$  & $2.5\%$ \\
    SFE$\in [0.05, 0.2]$ &  $\sum m_\nu < 0.111\,\mathrm{eV}$  & $8.3\%$  \\
    SFE$\in [0.05, 0.3]$ &  $\sum m_\nu < 0.111\,\mathrm{eV}$ & $8.3\%$  \\
    SFE$\in [0.05, 0.4]$ &  $\sum m_\nu < 0.114\,\mathrm{eV}$  & $5.8\%$ \\
    SFE$\in [0.05, 0.5]$ &  $\sum m_\nu < 0.117\,\mathrm{eV}$  & $3.3\%$ \\
    SFE$=0.1$ & $\sum m_\nu < 0.114\,\mathrm{eV}$  & $5.8\%$ \\
    SFE$=0.2$ & $\sum m_\nu < 0.110\,\mathrm{eV}$  & $9.1\%$ \\    
    SFE$=0.3$ & $\sum m_\nu < 0.113\,\mathrm{eV}$  & $6.6\%$ \\
    SFE$=0.4$ & $\sum m_\nu < 0.118\,\mathrm{eV}$  & $2.5\%$ \\        
    SFE$=0.5$ & $\sum m_\nu < 0.121\,\mathrm{eV}$  & $0$ \\
    \hline    
  \end{tabular}
\end{table}

By varying the prior on SFE, we also find an interesting phenomenon that, as Table~\ref{tab:SFEprior} shows, the best improvement of $\sum m_\nu$ bound is achieved around $\mathrm{SFE}\sim 0.2$. Physically this is because a very small SFE makes the likelihood approach a cosmology-independent constant $\mathrm{P}\rightarrow \mathrm{P}(N_{\rm obs} = 0)$, that all candidate galaxies are interpreted as false detections due to systematic errors in redshift and mass, and a very large SFE makes the likelihood approach another cosmology-independent constant ($\mathrm{P}\rightarrow 1$), that theoretical prediction significantly exceeds the observed number of galaxies.

\section{Conclusions and Discussion} \label{sect:conc}

In this work, we incorporate massive neutrinos into the numerical tool in \citealt{wang2023quantifying} to evaluate the likelihood of massive galaxy counting. At the current stage, the JWST candidate galaxies alone cannot constrain cosmology. However, in joint analyses with Planck or Planck+BAO, the addition of JWST massive galaxy counting improves the upper bound of neutrino masses by a few percent and puts the inverted mass ordering under more pressure.

While the Planck+BAO+JWST constraint gives the most stringent upper bound on neutrino masses, the results are yet very model-dependent. In particular, our results rely on  SFE prior at $z\sim 7$ , namely $\mathrm{SFE}\lesssim 0.3$. If a larger $\mathrm{SFE}\gtrsim 0.4$, which is unlikely in the usual scenario of galaxy formation and evolution,  is assumed, the likelihood becomes insensitive to cosmology and can no longer improve the $\sum m_\nu$ constraint. There are, however, alternative scenarios where an enhanced SFE at high redshift does not accelerate the reionization process, and therefore is consistent with observations. For example, \citealt{lin2023implications} considered large SFE ($\gtrsim 0.4$) at high redshift and warm dark matter in the $\mathrm{keV}$ mass range. In this picture, the warm dark matter has no impact on the abundance of very massive halos, and only suppresses the low-mass end of halo mass function. In other words, the $\mathrm{keV}$ warm dark matter reduces the number of low-mass galaxies that emit UV photons. This effect cancels the enhancement of UV radiation from each galaxy due to a large SFE,  and keeps the cosmic reionization history in line with observations. There are also other options, such as fuzzy dark matter, can suppress the abundance of small halos and galaxies and lead to the same phenomenon~\citep{gong2023fuzzy}.

Other cosmological probes, the type Ia supernovae, local distance-ladder measurements of $H_0$, weak gravitational lensing, redshift-space distortion, etc., may also tighten the current upper bound of neutrino masses. See e.g.~\citealt{aghanim2020planck} and discussion therein. Ideally, one would like to join all the available data and pushes the bound $\sum m_\nu$ to the best limit. However, at the current stage, these cosmological probes, including the JWST high-$z$ massive candidate galaxies, are not all in concordance and may suffer from unknown systematics~\citep{riess2021cosmic, asgari2021kids, huang2022s, wang2023revisiting}. It is thus not so favorable to add too many cosmological data into the pool at once, and caution should be taken for the cosmological bounds on neutrino masses. Nevertheless, the growing JWST data bring a hope to cross check the constraints on neutrino masses in very different perspectives (tail statistics, perturbation-based, nonlinear physics).

\normalem
\begin{acknowledgements}

This work is supported by National SKA Program of China No. 2020SKA0110402, the National Natural Science Foundation of China (NSFC) under Grant No. 12073088, and National key R\&D  Program of China (Grant No. 2020YFC2201600).

\end{acknowledgements}
  

\begin{thebibliography}{26}
\providecommand\natexlab[1]{#1}
\providecommand\JournalTitle[1]{#1}

\bibitem[{Abe} {et~al.}(2014)]{abe2014observation}
{Abe}, K., {Adam}, J., {Aihara}, H., {et~al.} 2014, \prl, 112, 061802

\bibitem[Aghanim {et~al.}(2020)]{aghanim2020planck}
Aghanim, N., Akrami, Y., Ashdown, M., {et~al.} 2020, Astronomy \& Astrophysics,
  641, A6

\bibitem[Asgari {et~al.}(2021)]{asgari2021kids}
Asgari, M., Lin, C.-A., Joachimi, B., {et~al.} 2021, Astronomy \& Astrophysics,
  645, A104

\bibitem[Boylan-Kolchin(2023)]{boylan2023stress}
Boylan-Kolchin, M. 2023, Nature Astronomy, 1

\bibitem[{Fukuda} {et~al.}(1998)]{fukuda1998evidence}
{Fukuda}, Y., {Hayakawa}, T., {Ichihara}, E., {et~al.} 1998, \prl, 81, 1562

\bibitem[Gong {et~al.}(2023)]{gong2023fuzzy}
Gong, Y., Yue, B., Cao, Y., \& Chen, X. 2023, The Astrophysical Journal, 947,
  28

\bibitem[{Goto} {et~al.}(2021)]{goto2021silverrush}
{Goto}, H., {Shimasaku}, K., {Yamanaka}, S., {et~al.} 2021, \apj, 923, 229

\bibitem[Haslbauer {et~al.}(2022)]{haslbauer2022has}
Haslbauer, M., Kroupa, P., Zonoozi, A.~H., \& Haghi, H. 2022, The Astrophysical
  Journal Letters, 939, L31

\bibitem[Howlett {et~al.}(2012)]{howlett2012cmb}
Howlett, C., Lewis, A., Hall, A., \& Challinor, A. 2012, \jcap, 1204, 027

\bibitem[Huang {et~al.}(2022)]{huang2022s}
Huang, L., Huang, Z., Zhou, H., \& Li, Z. 2022, Science China Physics,
  Mechanics \& Astronomy, 65, 239512

\bibitem[Labb{\'e} {et~al.}(2023)]{labbe2023population}
Labb{\'e}, I., van Dokkum, P., Nelson, E., {et~al.} 2023, Nature, 616, 266

\bibitem[Lewis(2019)]{lewis2019getdist}
Lewis, A. 2019, GetDist: a Python package for analysing Monte Carlo samples,
  arXiv:1910.13970

\bibitem[Lewis {et~al.}(2000)]{lewis2000efficient}
Lewis, A., Challinor, A., \& Lasenby, A. 2000, \apj, 538, 473

\bibitem[Lin {et~al.}(2023)]{lin2023implications}
Lin, H., Gong, Y., Yue, B., \& Chen, X. 2023, Implications of the Stellar Mass
  Density of High-$z$ Massive Galaxies from JWST on Warm Dark Matter,
  arXiv:2306.05648

\bibitem[Lovell {et~al.}(2023)]{lovell2023extreme}
Lovell, C.~C., Harrison, I., Harikane, Y., Tacchella, S., \& Wilkins, S.~M.
  2023, Monthly Notices of the Royal Astronomical Society, 518, 2511

\bibitem[Press \& Schechter(1974)]{press1974formation}
Press, W.~H., \& Schechter, P. 1974, The Astrophysical Journal, 187, 425

\bibitem[Qin {et~al.}(2023)]{qin2023implications}
Qin, Y., Balu, S., \& Wyithe, J. S.~B. 2023, Monthly Notices of the Royal
  Astronomical Society, 526, 1324

\bibitem[Riess {et~al.}(2021)]{riess2021cosmic}
Riess, A.~G., Casertano, S., Yuan, W., {et~al.} 2021, The Astrophysical Journal
  Letters, 908, L6

\bibitem[Sheth {et~al.}(2001)]{sheth2001ellipsoidal}
Sheth, R.~K., Mo, H., \& Tormen, G. 2001, Monthly Notices of the Royal
  Astronomical Society, 323, 1

\bibitem[Sheth \& Tormen(1999)]{sheth1999large}
Sheth, R.~K., \& Tormen, G. 1999, Monthly Notices of the Royal Astronomical
  Society, 308, 119

\bibitem[Sheth \& Tormen(2002)]{sheth2002excursion}
Sheth, R.~K., \& Tormen, G. 2002, Monthly Notices of the Royal Astronomical
  Society, 329, 61

\bibitem[Trenti \& Stiavelli(2008)]{trenti2008cosmic}
Trenti, M., \& Stiavelli, M. 2008, The Astrophysical Journal, 676, 767

\bibitem[{Wang} {et~al.}(2023{\natexlab{b}})]{wang2023revisiting}
{Wang}, J., {Huang}, Z., \& {Huang}, L. 2023{\natexlab{b}}, Science China
  Physics, Mechanics, and Astronomy, 66, 129511

\bibitem[{Wang} {et~al.}(2023{\natexlab{a}})]{wang2023quantifying}
{Wang}, J., {Huang}, Z., {Huang}, L., \& {Liu}, J. 2023{\natexlab{a}}, arXiv
  e-prints, arXiv:2311.02866

\bibitem[{Wold} {et~al.}(2022)]{wold2022large}
{Wold}, I. G.~B., {Malhotra}, S., {Rhoads}, J., {et~al.} 2022, \apj, 927, 36

\bibitem[{Zhang} {et~al.}(2022)]{zhang2022massive}
{Zhang}, Z., {Wang}, H., {Luo}, W., {et~al.} 2022, \aap, 663, A85

\end{thebibliography}

\end{document}